%% file: HIG-11-028_temp.tex
\pdfoutput=1

\documentclass[11pt,twoside,a4paper,cmspaper,final,collab]{cms-tdr}
\def\svnVersion{104476:104533}\def\cmsCernNo{2012-038}\def\cmsCernDate{\today}\def\cmsMessage{Submitted to the Journal of High Energy Physics}

\begin{document}\cmsNoteHeader{HIG-11-028}

\hyphenation{had-ron-i-za-tion}
\hyphenation{cal-or-i-me-ter}
\hyphenation{de-vices}

\RCS$Revision: 104474 $
\RCS$HeadURL: svn+ssh://alverson@svn.cern.ch/reps/tdr2/papers/HIG-11-028/trunk/HIG-11-028.tex $
\RCS$Id: HIG-11-028.tex 104474 2012-02-16 11:52:04Z asavin $

\newcommand{\tauh}{\ensuremath{\Pgt_\mathrm{h}}}
\newcommand{\taumu}{\ensuremath{\Pgt_{\Pgm}}}
\newcommand{\taue}{\ensuremath{\Pgt_{\Pe}}}
\newcommand{\taus}{\ensuremath{\ell^{+}\ell^{-}\Pgt^{+}\Pgt^{-}}}
\newcommand{\PZ}{\cPZ}
\providecommand{\mH}{\ensuremath{m_{\PH}\xspace}}
\cmsNoteHeader{HIG-11-028} % This is over-written in the CMS environment: useful as preprint no. for export versions
\title{Search for the standard model Higgs boson in the $\PH \rightarrow \PZ\PZ \rightarrow \ell^+\ell^-\Pgt^+\Pgt^-$ decay channel in \Pp\Pp\ collisions at $\sqrt{s}=7$\TeV}

\date{\today}

\abstract{
A search is reported for the standard model Higgs boson in the
$ \PH \to \PZ\PZ \to \taus$ decay mode, where $\ell = \mu$ or $\Pe$, in
proton-proton collisions at $\sqrt{s}=7$\TeV,
corresponding to an integrated
luminosity of 4.7\fbinv collected with the
CMS detector at the LHC.
No evidence is found for a significant deviation from the
background expectation.
An upper limit four to twelve  times larger than the predicted value
is set at 95\% confidence level for
the product of the standard model
Higgs boson production cross section and decay branching fraction
in the mass range $190 < \mH < 600\GeV$.
}

\hypersetup{%
pdfauthor={CMS Collaboration},%
pdftitle={Search for the standard model Higgs boson in the H to ZZ to ll tau tau decay channel in pp collisions at sqrts=7 TeV},
pdfsubject={CMS},%
pdfkeywords={CMS, physics, Higgs}}

\maketitle %maketitle comes after all the front information has been supplied

\section{Introduction}

The search for the
standard model (SM)~\cite{Glashow:1961tr,Weinberg:1967tq,sm_salam}
Higgs boson is
one of the main goals of the Large Hadron Collider (LHC) physics programme.
The discovery of the SM Higgs boson
 would shed light on the spontaneous electroweak symmetry breaking
 mechanism~\cite{Englert:1964et,Higgs:1964ia,Higgs:1964pj,Guralnik:1964eu,Higgs:1966ev,Kibble:1967sv}.
To date, experimental searches for this particle have yielded null results.
Limits at 95\% confidence level (CL) on its mass have been placed by experiments
at LEP,  $\mH > 114.4$\GeV~\cite{LEP},  the Tevatron,
$\mH \notin$~(162--166)\GeV~\cite{TEVHIGGS_2010},
and ATLAS, $\mH \notin$~(145--206), (214--224),
(340--450)\GeV~\cite{ATLAS:2011aa, Aad:2011uq, ATLAS:2011af}.
Precision electroweak measurements, not taking into account the results from direct searches,
indirectly constrain the SM Higgs boson mass to be less than 158\GeV~\cite{EWKlimits}.

This letter presents a  search for the SM Higgs boson in the decay
 mode $ \PH \to \PZ\PZ \to \taus$, where $\ell$ is either $\Pgm$ or $\Pe$.
One $\PZ$ is required to decay either
into $\Pgmp\Pgmm$ or $\Pep\Pem$, and the second
$\PZ$ into $\Pgt^+\Pgt^-$ in four possible final states: $\tauh \tauh$, $\taumu \tauh$, $\taue \tauh$, and
$\taue \taumu$,
where $\tauh$ represents a $\Pgt$  decaying
hadronically, and $\taumu$, and $\taue$ indicate taus decaying into muons and electrons respectively.
The present measurement complements the
search in the $\PH \to \PZ\PZ \to 4\ell $ channel~\cite{CMS_4l}.
The presence of four leptons in the final state provides a clean signature with only a small
contribution from background processes.
The major irreducible background contribution
is non-resonant  $\PZ\PZ$ production. The most important reducible background contributions are
 $\PZ$ and $\PW\PZ$ production in association with jets, and $\ttbar$ production.
The final states $\Pgt\Pgt \to \taumu \taumu, \taue \taue$ are not considered, as
they are accounted for in the $\PH \to \PZ\PZ \to 4\ell$  Higgs search~\cite{CMS_4l}.
The cross sections for the Higgs boson production mechanisms and decay branching fractions,
together with their uncertainties, are taken from
Ref.~\cite{LHCHiggsCrossSectionWorkingGroup:2011ti}
and are derived from
Refs.~\cite{
      Djouadi:1991tka,
      Dawson:1990zj,Spira:1995rr,
      Djouadi:1997yw,
      Harlander:2002wh,Anastasiou:2002yz,Ravindran:2003um,
      Catani:2003zt,
      Figy:2003nv,
      Aglietti:2004nj,Degrassi:2004mx,
      Bredenstein:2006rh,Bredenstein:2006ha,
      Ciccolini:2007jr,Ciccolini:2007ec,
      Actis:2008ug,
      Anastasiou:2008tj,deFlorian:2009hc,
      Bolzoni:2010xr,
      Baglio:2010ae}.

The analysis is
based on
 data from proton-proton collisions at $\sqrt{s}=7$\TeV,
corresponding to an integrated
luminosity of 4.7\fbinv collected with the
Compact Muon Solenoid (CMS) detector at the LHC in 2011.
This is the first Higgs boson search performed in the  $\PH \to \PZ\PZ \to \taus$ channel.

\section{CMS Detector}

A detailed description of the CMS detector can be found elsewhere~\cite{CMSExperiment}.
The central feature of the CMS apparatus
is a superconducting solenoid, of 6\unit{m} internal diameter, providing
a magnetic field of 3.8\unit{T}. Within the field volume are the silicon pixel
and strip tracker, the crystal electromagnetic calorimeter (ECAL),
and the brass/scintil\-lator hadron calorimeter.
Muons are measured in gas-ionization detectors embedded
in the steel return yoke.

CMS uses a right-handed coordinate system, with the
origin at the nominal interaction point, the $x$-axis
pointing to the centre of the LHC ring, the $y$-axis pointing
up (perpendicular to the LHC plane), and the $z$-axis
along the counterclockwise-beam direction. The polar angle,
$\theta$, is measured from the positive $z$-axis and the
azimuthal angle, $\phi$, is measured in the $x$-$y$ plane.
Variables used in this analysis are
the pseudorapidity,
$\eta = -\ln[\tan(\theta/2)]$, and the transverse
momentum, $\PT = \sqrt{p_x^2 + p_y^2}$.

The ECAL is designed to have both excellent energy resolution
and high granularity, properties that are crucial for reconstructing
electrons and photons produced in $\Pgt$-lepton decays.
The ECAL is constructed with projective lead tungstate
crystals that
provide coverage in pseudorapidity $\vert \eta \vert< 1.479 $ in a barrel region and
$1.479 <\vert \eta \vert < 3.0$ in two endcap regions (EE). A preshower detector consisting of two planes
of silicon sensors interleaved with a total of 3$\,X_0$ of lead is located in front of the EE.
The energy resolution is 3\% or better for the range of electron energies relevant for this analysis.

The inner tracker measures charged particle tracks
within the range $|\eta| < 2.5$.  It consists of
$1\, 440$ silicon pixel and $15\, 148$ silicon strip detector modules,
and provides an impact parameter resolution of
${\sim}15\micron$ and a transverse momentum
resolution of about 1.5\% for
100\GeV particles.
The reconstructed tracks are used to measure the location of interaction vertices.
The spatial resolution of the reconstruction is
${\approx}25\micron$ for vertices with more than 30 associated tracks~\cite{tracker}.

The muon barrel region is covered
by drift tubes, and the endcap regions by cathode
strip chambers.  In both regions, resistive plate
chambers provide additional coordinate and timing
information.  Muons are reconstructed in the
range $|\eta| < 2.4$, with a typical $\PT$
resolution of ${\approx}1\%$ for $\PT \approx 40$\GeV.

\section{Event Selection and
Monte Carlo Samples}

At the trigger level, the selected events are
required to have at least two muons, one
with $\pt > 13\GeV$ ($\pt > 17\GeV$ for the end of the data-taking period
when the instantaneous luminosity was highest)
 and the other with $\pt > 8\GeV$,
or at least two electrons, one
with $\pt > 17\GeV$ and the other with $\pt > 8\GeV$.

Algorithms for identifying muons and electrons, collectively referred to as leptons, are
 based on the tracker, the muon systems and the calorimeters
 ~\cite{CMS-PAS-MUO-10-002,CMS-PAS-EGM-10-004}.
Since the $\PZ\PZ$ final state is expected to
have only a small contribution from background processes, the algorithms are
tuned to maximize the lepton-reconstruction efficiency,
resulting in an increased lepton-misidentification rate.
A particle flow (PF) technique~\cite{CMS-PAS-PFT-09-001} is used
to form lepton-isolation quantities and is also used for
$\tauh$ reconstruction.
In the
PF approach, information from all subdetectors is combined
to reconstruct and identify particles produced in the collision.
The particles are classified into mutually exclusive
categories:
charged hadrons, photons, neutral hadrons, muons, and
electrons.
These particles are used to reconstruct
$\tauh$
with
the ``hadron plus strip'' (HPS) algorithm~\cite{CMS-PAS-TAU-11-001} that
is designed to optimize the performance of
$\tauh$ identification and reconstruction
by considering specific $\tauh$
decay modes.
The neutrinos produced in all $\Pgt$ decays escape detection
and are ignored in the $\tauh$ reconstruction.
The algorithm provides high $\tauh$ identification efficiency,
approximately 50\% for the range of $\tauh$ energies relevant for this analysis,
while keeping the misidentification rate for jets at the level
of ${\approx}1\%$, that is factor of three to four times lower
with respect to other available
algorithms~\cite{PTDR}.

Events are required to have at least one
$\PZ \to \ell^+\ell^-$ candidate, denoted by $\PZ_1$,
with the leptons of opposite charge, one
with
$\pt > 20\GeV$ and another with $\pt > 10\GeV$,
and with
$|\eta| <  2.4$ for the muons and
$|\eta| < 2.5$ for the electrons.
Both leptons are required to have
a combined PF relative isolation $I_\text{rel}^\mathrm{PF} < 0.25$, which is defined as:
\begin{equation}
 I_\text{rel}^\mathrm{PF} = {\left (\pt^\text{charged}+\text{max}(\ET^{\Pgg} + \ET^\text{neutral}-0.5\times \PT^\mathrm{PU},0)
 \right )}/{\pt^\ell},
\end{equation}
where $\pt^{\text{charged}}$ is the scalar sum of the
charged hadrons \pt,
and
$\ET^{\Pgg}$ and $\ET^{\text{neutral}}$ correspond, respectively,
to the sum of the transverse energies of the
photons and neutral hadrons,
all measured in
the isolation cone of
 $\Delta R < 0.4$
around the lepton direction, where
$\Delta R = \sqrt { (\Delta \eta)^2 + (\Delta \phi)^2 }$.
The contribution from neutrals is corrected for the effect of overlapping
pp interactions.
The value of the correction is estimated by scaling
the sum of the $\PT$ of all charged particles not associated with the vertex
in the isolation cone
($\PT^\mathrm{PU}$)
by  a factor 0.5, which is
 approximately the ratio of neutral to charged hadron yields in
hadronization processes.

For the second $\PZ$, denoted by $\PZ_2$, the selection requirements
depend
on the final state. If the final state is $\taumu \taue$,
the lepton $\pt$ values are required
to exceed 10\GeV. The remaining criteria are
identical to those
for $\PZ_1$.
Since $\tauh$s have much larger misidentification rates
than the other leptons,
the isolation requirement based on $I_\text{rel}^\mathrm{PF}$ for the muons and
electrons in the final states $\taumu \tauh$ and $\taue \tauh$ is changed to
0.15 and 0.1, respectively.
In a study of inclusive $\PZ \to \tau\tau$ production~\cite{CMS-EWK-TAU}, it was demonstrated
that modifying the muon and electron
isolation requirements is  a more effective way to reduce background in such final
states than
requiring tighter isolation on $\tauh$.
The $\tauh$ are required to have $\pt > 20\GeV$ and $|\eta| <  2.3$, and to pass a loose
HPS working-point requirement. If the $\PZ_2$ decays to $\tauh^+ \tauh^-$,
both $\tauh$ are required to pass a medium working point of the HPS algorithm.
The loose (medium) working point requires the scalar sum over the
charged
hadrons {\pt} and the neutral hadrons {\ET} in the isolation cone,
to be less
than 2\GeV (1\GeV).
The loose (medium) working point corresponds to a probability of approximately
1\% (0.5\%) for jets to be misidentified as $\tauh$. Using the medium
instead of loose
working
point
leads to a decrease in the
$\tauh$ reconstruction efficiency
from
${\approx}50\%$ to ${\approx}40\%$.

The visible invariant mass of the
reconstructed $\PZ_2 \to \Pgt^+\Pgt^-$ is required to be
$ 30 < m_{\Pgt\Pgt} < 80\GeV$, and this criterion is used for most of the final states.
The upper bound
 reduces contributions
from $\PZ_2 \to \ell\ell$, where a muon or an electron is not well reconstructed, and misidentified as a $\tauh$.
For the $\PZ_2 \to \taue\taumu$ final state, the upper bound is increased to 90\GeV, as this
state is not produced in $\PZ_2 \to \ell\ell$ decays.
Leptons from the same $\PZ$ are required to be
separated by $\Delta R > 0.4$ for $\PZ_1$, and
by $\Delta R > 0.5$ for $\PZ_2$.
The two reconstructed $\PZ_1$ and $\PZ_2$ are required to be
separated by $\Delta R > 0.5$.

A set of Monte Carlo (MC) event samples is used
to simulate signal and
background events.
The
Drell--Yan background, $\ell^+\ell^-$ in association with jets, is simulated with
the next-to-leading order (NLO) MC generator
\POWHEG2.0~\cite{Alioli:2008gx,Nason:2004rx,Frixione:2007vw}.
The QCD multijet, $\PW$ and diboson $\PW\PZ$
backgrounds are simulated with \textsc{pythia 6.424}~\cite{Sjostrand:2006za}.
The $\PZ\PZ$ background is simulated
with \PYTHIA6.424 and
\MADGRAPH~\cite{Maltoni:2002qb}.
The $\ttbar$ samples  are simulated with
\MADGRAPH.
The $\Pgt$--lepton decays are generated with \textsc{tauola}~\cite{Jadach:1993hs}.
The Higgs boson samples are generated using \POWHEG2.0, which
incorporates NLO gluon fusion ($\Pg\Pg \to \PH$) and vector-boson fusion ($\cPq\cPq \to \cPq\cPq\PH$).
All events are processed through a detailed simulation of the CMS detector based on
\GEANTfour~\cite{GEANT} and reconstructed with the same algorithms that are used for data.

\section{Background Estimates and
Systematic Uncertainties}

The major irreducible source of background
to the $\PH \to \PZ\PZ \to \taus$ process is
from SM
$\PZ\PZ \to \taus$ production. The $\PZ\PZ$ contribution
is estimated from data by scaling the
prediction from simulation to the well measured inclusive $\PZ$
production cross section.
The number of estimated $\PZ\PZ$ events, $N_{\PZ\PZ}^\text{est}$, can be written as:
\begin{equation}
N_{\PZ\PZ}^\text{est} = N_\PZ^\text{obs} \cdot \frac {\sigma_{\PZ\PZ}^\mathrm{SM}}{\sigma_\PZ^\mathrm{SM}} \cdot \frac {A_{\PZ\PZ}}{A_\PZ},
\end{equation}
where $N_\PZ^\text{obs}$ is the number of observed events from
inclusive $\PZ$ production,
$A_\PZ$ is their estimated acceptance
from a MC simulation,
including all selection requirements,
and rescaled by measured data/MC correction factors,
$A_{\PZ\PZ}$ is the acceptance for $\PZ\PZ$ events,
$\sigma_\PZ^\mathrm{SM}$ is the
SM cross section for inclusive $\PZ$ production, and
$\sigma_{\PZ\PZ}^{\rm SM}$ is the SM cross section for $\PZ\PZ$ production calculated
with MCFM~\cite{MCFM}.

The other major background contributions arise from
the production of $\PZ$ and $\PW\PZ$ in association with
jets, as well as
$\ttbar$ and
QCD multijet production. The latter two backgrounds are small.
In all these cases, a
jet or non-isolated lepton is misidentified as a $\tauh$, $\taue$ or $\taumu$.
The probability for jets to be misidentified as $\tauh$ is measured using
$\ell^+\ell^-\tauh\tauh$ events in data in which the
$\PZ_1$ passes all selection requirements,
but no requirement is applied on $\tauh$ isolation, and
 the two $\tauh$ candidates are required to have the same charge.
This region is dominated by $\PZ$+jets events.
The $\tauh$ misidentification rate is defined as the ratio of the number of $\tauh$ candidates that pass
the HPS loose or medium working-point requirements,
to the initial number of $\tauh$ candidates, and  is measured as a function of the
$\pt$ for each $\tauh$.
To estimate the number of background events in the signal region, the
measured misidentification rate
is applied
to  events that  pass all selection requirements, including
the opposite-charge requirement for the $\PZ_2$,
but requiring
the $\tauh$ candidates to not be isolated.

The misidentification rate for $\taue$ and $\taumu$ in the $\Pgm\Pgm\taumu \taue$ and
$ee \taumu \taue$ final states is estimated using events in which the
$\PZ_1$ passes all selection requirements, and the event contains
an additional muon or electron. No isolation
requirements are applied to it. The misidentification rate is
defined for $\taue$ and $\taumu$
in the same way
as described above for $\tauh$  and applied to $\Pgm\Pgm\taumu \taue$ and $ee \taumu \taue$ events  that
pass all the selection requirements, but requiring $\taue$ or $\taumu$ to not be isolated.
Isolated muons and electrons from
$\PH \to \PZ\PZ \to 4\ell$ and
$\PZ\PZ \to 4\ell$  production can also be misidentified as $\tauh$.
Events are rejected
if they are also identified as $ \PZ\PZ \to 4\ell$ events with criteria described in
Ref.~\cite{CMS_4l}.

Theoretical uncertainties on the Higgs boson cross section (17--20\%) and
branching ratio (2\%) are taken from Ref.~\cite{LHCHiggsCrossSectionWorkingGroup:2011ti}.
Recent studies \cite{Passarino:2010qk,LHCHiggsCrossSectionWorkingGroup:2011ti,Anastasiou:2011pi}
show that current MC simulations do not describe the correct Higgs boson mass line shape
above ${\approx}300\GeV$.
This effect amounts to an additional uncertainty on the theoretical cross section,
and hence on the limits, of about 4\% at $\mH = 300\GeV$ and
10--30\% for $\mH$ of 400--600\GeV.
The main uncertainty on the estimate of the $\PZ\PZ$ background arises from the
theoretical uncertainty on the $\PZ\PZ$ production cross section and is taken from Ref.~\cite{LHChiggs}.
The uncertainties on the other backgrounds, \PZ+jets, \PW\PZ+jets, and $\ttbar$ reflect
the uncertainties on the measured values of the misidentication rates and the
limited statistics of the control regions in the data.
The uncertainty on integrated luminosity of the data sample is 4.5\%~\cite{lumiPAS1}.
Systematic uncertainties on trigger efficiency (1\%) and on lepton
identification efficiency and isolation are evaluated from data.
The uncertainties associated with lepton
identification and isolation are 1-2\% for muons and electrons, and 6-7\%
for $\tauh$.
Uncertainties on energy scales, 3\% for $\tauh$ and 1-2.5\% for electrons, contribute
to variation in the shape of the mass spectrum.

\section{Results}

\begin{figure}[htbp]
\begin{center}
\includegraphics[scale=0.50]{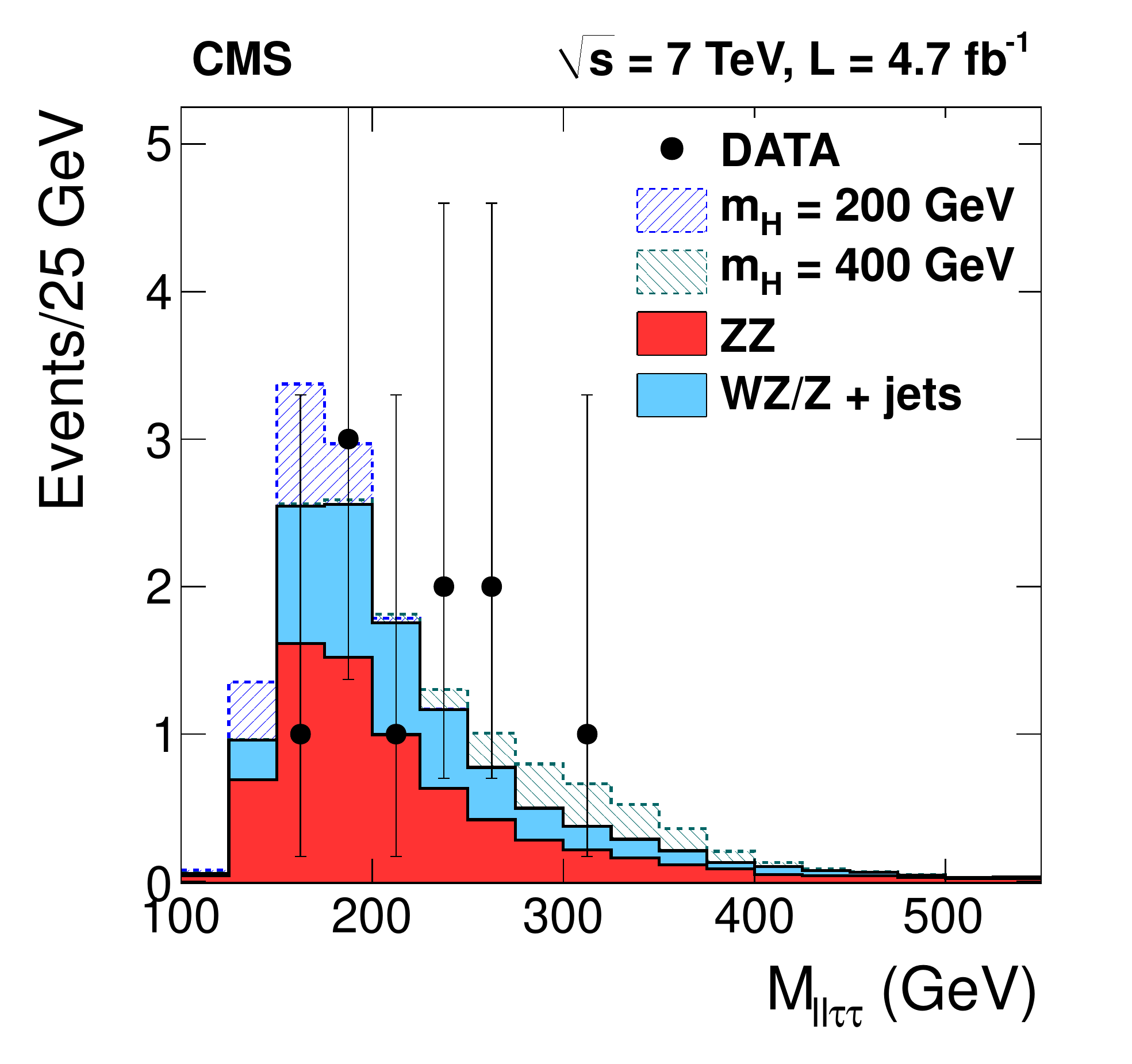}\\
\caption{The four-lepton reconstructed mass summed for all $\taus$ final states.
The data corresponds to an integrated luminosity of
4.7 \fbinv. Points represent the data, shaded histograms represent
the background
and hashed histograms represent the signal expectations for two Higgs boson
masses.
The background shapes are taken from MC simulation and are normalized
to the values obtained using control data samples, as
described in the text.
}
\label{figure:invmass}
\end{center}
\end{figure}

Ten $\taus$ candidates are observed
in eight search channels, while
 $11.60 \pm 0.54\stat \pm 1.62\syst$
background events are expected.
Table~\ref{table:results} compares
the estimated number of background events to the number of events observed
in the signal region.
The distribution of the reconstructed invariant mass summed over  all eight
 $\taus$ decay channels is shown in
Fig.~\ref{figure:invmass}.
The shape of the background is taken from the MC simulation, with each
component normalized to
the corresponding estimated value from Table~\ref{table:results}.
The expected mass distributions for the SM Higgs boson with a mass of
$\mH = 200\GeV$ and $400\GeV$ are also shown in Fig.~\ref{figure:invmass}.
The reconstructed masses are shifted with respect to the generated values by $\approx 30\%$ due to the
undetected  neutrinos in $\Pgt$ decays. As a result, the
$\PH \to \PZ\PZ \to \taus$ mass resolution is 10-15\%,
depending on the final state, and is almost independent of $\mH$.

\begin{table}[htbp]
\centering
\caption{ The estimated yields of $\PZ\PZ$ and other background
events obtained from data,
as described in the text, are shown for each decay channel and
are summed in the total background yield (``Total backgr.''),
and compared to the number of events observed in the signal region.
The total uncertainty is the sum in quadrature of
statistical and systematic uncertainties.
The number of signal
events expected for the SM Higgs boson
with a mass of $\mH = 200\GeV$ is also shown.}
\label{table:results}
\begin{tabular}{|c|c|c|c|c|c|}
\hline
Decay & $N_{\PZ\PZ}^{est}$ &  Other   & Total & $\mH$ & Observed \\
channel &  &  backgrounds  & backgr. & $200\GeV$ & \\
\hline
\hline
$\Pgm\Pgm\tauh\tauh$ 			& $ 0.79 \pm 0.09 $ & $ 0.76 \pm 0.31 $ & $1.55 \pm 0.32$ & 0.17 & $ 0 $\\
$\Pe\Pe\tauh\tauh$ 				& $ 0.75 \pm 0.09 $ & $ 0.73 \pm 0.32 $ & $1.48 \pm 0.33$ & 0.15 & $ 1 $\\
$\Pe\Pe\Pgt_{\Pe}\tauh$ 			& $ 1.12 \pm 0.13 $ & $ 0.99 \pm 0.34 $ & $2.11 \pm 0.36$ & 0.25 & $ 3 $\\
$\Pgm\Pgm\Pgt_{\Pe}\tauh$ 		& $ 1.20 \pm 0.14 $ & $ 0.31 \pm 0.29 $ & $1.51 \pm 0.32$ & 0.26 & $ 3 $\\
$\Pgm\Pgm\Pgt_{\Pgm}\tauh$ 	& $ 1.08 \pm 0.13 $ & $ 0.67 \pm 0.36 $ & $1.75 \pm 0.38$ & 0.23 & $ 2 $\\
$\Pe\Pe\Pgt_{\Pgm}\tauh$ 		& $ 0.94 \pm 0.10 $ & $ 0.41 \pm 0.16 $ & $1.35 \pm 0.19$ & 0.20 & $ 0 $\\
$\Pe\Pe\Pgt_{\Pe}\Pgt_{\Pgm}$ 		& $ 0.51 \pm 0.06 $ & $ 0.58 \pm 0.42 $ & $1.09 \pm 0.42$ & 0.11 & $ 0 $\\
$\Pgm\Pgm\Pgt_{\Pe}\Pgt_{\Pgm}$ 	& $ 0.58 \pm 0.07 $ & $ 0.18 \pm 0.18 $ & $0.76 \pm 0.22$ & 0.12 & $ 1 $\\
\hline
\hline
Total & $ 6.97 \pm 0.84 $ & $4.63 \pm 1.49 $ & $11.60 \pm 1.71$ & 1.49 & 10 \\
\hline
\end{tabular}
\end{table}

The product of the acceptance and branching fraction for
the individual $\tau$-decay channels ranges
between 0.01--0.02 for $\mH = 200\GeV$, and increases by a factor of
three to four for
$\mH = 400$--450\GeV. This behaviour is expected. The
final-state leptons produced in $\Pgt$
decays of more massive Higgs bosons have higher momenta than those from
direct $\PZ \to \ell\ell$ production and the selection requirements and the
lepton reconstruction  become more efficient
at larger $\mH$.
While the cross section decreases with increasing $\mH$,
the expected number of Higgs boson events selected by this analysis in 4.7\fbinv of pp
collisions at 7\TeV is 1.4--1.5
in the range  $200 < \mH < 400\GeV$, and
decreases rapidly at higher masses.

\begin{figure}[htbp]
\begin{center}
\includegraphics[scale=0.50]{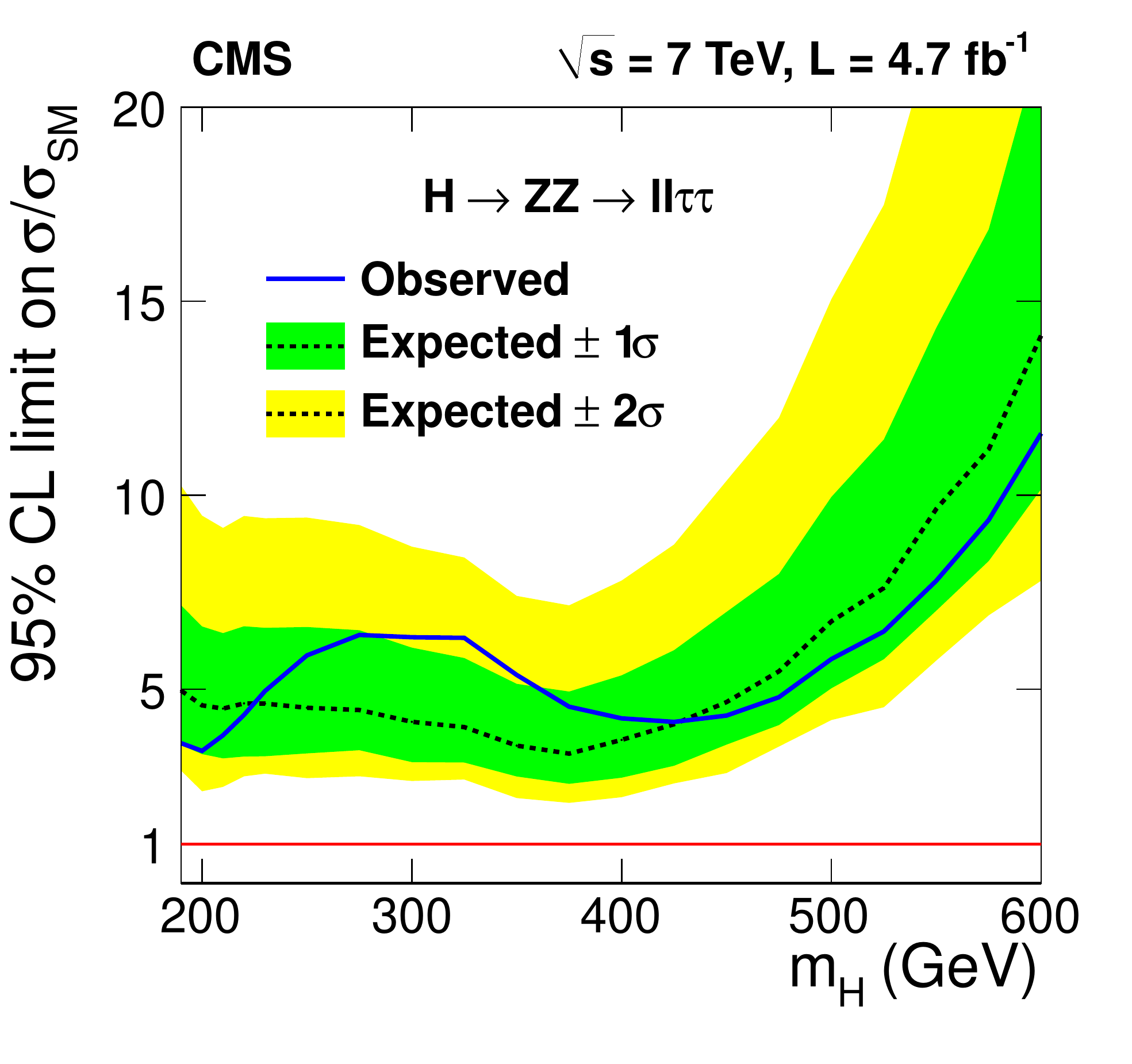}\\
\caption{The expected and observed  95\% CL upper limits on
the product of the Higgs boson production cross section and decay branching fraction
normalized to the SM expectation,
denoted by $\sigma / \sigma_{\rm{SM}}$,
as a function of $\mH$.
The one- and two-standard-deviation variations from the expected limit are also
 shown by the green and yellow bands.}
\label{figure:limitPlot}
\end{center}
\end{figure}

In Fig.~\ref{figure:limitPlot} the
expected and observed upper limits at 95\% CL on 
the product of the Higgs boson production cross section and decay branching fraction
normalized to the SM expectation
are presented
 as a function of $\mH$.
The limits are calculated with the modified frequentist
construction CL$_{\mathrm{s}}$~\cite{CLS1, CLS2, ATL-PHYS-PUB-2011-011} based on the shape of the
$\taus$ invariant mass distributions by including all eight individual
channels in the likelihood
combination.
The green and yellow bands represent the one- and two-standard-deviation variations from the expected limit.
The systematic uncertainties are introduced in the form of nuisance
parameters with log-normal probability density functions.
The upper limit on the cross section is approximately a factor four to twelve larger than
 the SM Higgs boson production cross section in the range of
$190 < \mH < 600\GeV$.

\section{Summary}

A search
for the standard model Higgs boson
has been performed
in the decay mode $ \PH \to \PZ\PZ \to \taus$
using CMS data corresponding to an integrated luminosity of 4.7\fbinv.
No evidence is found for a significant deviation from the background expectation.
An upper limit four to twelve  times larger than the predicted value
is set at 95\% confidence level for
the product of the standard model
Higgs boson production cross section and decay branching fraction
in the mass range $190 < \mH < 600\GeV$.
This is the first Higgs boson search performed in the  $\PH \to \PZ\PZ \to \taus$ channel.

\section*{Acknowledgements}

We congratulate our colleagues in the CERN accelerator departments for the excellent performance of the LHC machine. We thank the technical and administrative staff at CERN and other CMS institutes, and acknowledge support from: FMSR (Austria); FNRS and FWO (Belgium); CNPq, CAPES, FAPERJ, and FAPESP (Brazil); MES (Bulgaria); CERN; CAS, MoST, and NSFC (China); COLCIENCIAS (Colombia); MSES (Croatia); RPF (Cyprus); MoER, SF0690030s09 and ERDF (Estonia); Academy of Finland, MEC, and HIP (Finland); CEA and CNRS/IN2P3 (France); BMBF, DFG, and HGF (Germany); GSRT (Greece); OTKA and NKTH (Hungary); DAE and DST (India); IPM (Iran); SFI (Ireland); INFN (Italy); NRF and WCU (Korea); LAS (Lithuania); CINVESTAV, CONACYT, SEP, and UASLP-FAI (Mexico); MSI (New Zealand); PAEC (Pakistan); MSHE and NSC (Poland); FCT (Portugal); JINR (Armenia, Belarus, Georgia, Ukraine, Uzbekistan); MON, RosAtom, RAS and RFBR (Russia); MSTD (Serbia); MICINN and CPAN (Spain); Swiss Funding Agencies (Switzerland); NSC (Taipei); TUBITAK and TAEK (Turkey); STFC (United Kingdom); DOE and NSF (USA).

Individuals have received support from the Marie-Curie programme and the European Research Council (European Union); the Leventis Foundation; the A. P. Sloan Foundation; the Alexander von Humboldt Foundation; the Belgian Federal Science Policy Office; the Fonds pour la Formation \`a la Recherche dans l'Industrie et dans l'Agriculture (FRIA-Belgium); the Agentschap voor Innovatie door Wetenschap en Technologie (IWT-Belgium); the Council of Science and Industrial Research, India; and the HOMING PLUS programme of the Foundation for Polish Science, cofinanced from European Union, Regional Development Fund.

\bibliography{auto_generated}   % will be created by the tdr script.

\cleardoublepage \appendix\section{The CMS Collaboration \label{app:collab}}\begin{sloppypar}\hyphenpenalty=5000\widowpenalty=500\clubpenalty=5000\input{HIG-11-028-authorlist.tex}\end{sloppypar}
\end{document}

%% file: HIG-11-028-authorlist.tex
\textbf{Yerevan Physics Institute,  Yerevan,  Armenia}\\*[0pt]
S.~Chatrchyan, V.~Khachatryan, A.M.~Sirunyan, A.~Tumasyan
\vskip\cmsinstskip
\textbf{Institut f\"{u}r Hochenergiephysik der OeAW,  Wien,  Austria}\\*[0pt]
W.~Adam, T.~Bergauer, M.~Dragicevic, J.~Er\"{o}, C.~Fabjan, M.~Friedl, R.~Fr\"{u}hwirth, V.M.~Ghete, J.~Hammer\cmsAuthorMark{1}, M.~Hoch, N.~H\"{o}rmann, J.~Hrubec, M.~Jeitler, W.~Kiesenhofer, M.~Krammer, D.~Liko, I.~Mikulec, M.~Pernicka$^{\textrm{\dag}}$, B.~Rahbaran, C.~Rohringer, H.~Rohringer, R.~Sch\"{o}fbeck, J.~Strauss, A.~Taurok, F.~Teischinger, P.~Wagner, W.~Waltenberger, G.~Walzel, E.~Widl, C.-E.~Wulz
% ...
\vskip\cmsinstskip
\dag:~Deceased\\
1:~~Also at CERN, European Organization for Nuclear Research, Geneva, Switzerland\\
% ...
